\documentclass{kluwer}    

\usepackage{graphicx}

\newdisplay{guess}{Conjecture}

\begin{document}
\begin{article}
\begin{opening}
\title{Nonlinearity of nonradial modes in evolved stars
}
\author{Rafa{\l} M. \surname{Nowakowski}$^1$,
 Wojciech A. \surname{Dziembowski}$^{1,2}$}
\runningauthor{R.M. Nowakowski, W.A. Dziembowski}
\runningtitle{Nonlinearity of nonradial modes in evolved stars}
\institute{{1} Copernicus Astronomical Center,
  Bartycka 18, 00-716 Warsaw, Poland\\
 {2} Warsaw University Observatory, Al. Ujazdowskie 4, 00-478 Warsaw, Poland\\}
\date{August 28, 2002}

\begin{abstract}
We show that in evolved stars, even at relatively low surface
amplitudes, nonradial modes become strongly nonlinear in the
hydrogen shell source, where the Brunt-V\"ais\"al\"a frequency has
its absolute maximum. The measure of nonlinearity is the product
of horizontal displacement times the radial wavenumber, $|\xi_H
k_r|$. It becomes large already in evolved $\delta$-Scuti stars.
This nonlinearity presents a major problem for interpretations of
amplitude modulation in RR Lyrae stars in terms of nonradial mode
excitation.
\end{abstract}

\end{opening}

\section{Introduction}

All recent models of Blazhko-type modulation in RR Lyrae stars
postulate departures from pure radial pulsation. In the oblique
pulsator model \cite{Shibahashi} the $\ell=2$ component is induced
by magnetic field. In the resonant model \cite{Nowakowski}
low-$\ell$ modes are excited due to a resonant coupling with the
radial mode. Amplitudes of postulated nonradial components are quite
sizable, according to Kov\'acs \shortcite{Kovacs} they reach up to 0.7
radial mode amplitudes.

Apart of RR Lyrae stars, nonradial oscillations with quite high
amplitudes are observed in some evolved $\delta$-Scuti stars. The
best example is a post-MS star 4CVn, where several modes have been
identified as those of  $\ell=1$ or $2$ \cite{Breger}.

The nonradial modes observed in RR Lyrae and evolved
$\delta$-Scuti stars are of a mixed character. While it is known
that such modes at observed amplitudes are only weakly nonlinear
in the p-mode cavity, the nonlinearity in the g-mode cavity, where
most of the mode energy is concentrated, has never been studied
carefully.

\section{A measure of nonlinearity}

We study the nonlinearity of the g-waves in the asymptotic
approximation, which may be applied when the oscillation frequency
is much smaller than the Brunt-V\"ais\"al\"a and Lamb frequencies.
Then we have, approximately,
\begin{equation}
{\mbox{\boldmath
$\xi$}}=A(r)\left(\cos\Phi(r){\mbox{\boldmath{$e$}}}_r-\frac{r
k_r}{\sqrt{l(l+1)}} \sin\Phi(r)\mbox{\boldmath $\nabla$}\!_H\right)
Y_l^m(\theta,\phi)\exp(-{\rm i}\omega t), \label{xi}\\
\end{equation}
where \mbox{\boldmath $\xi$} is the displacement, $A(r)$ is a
slowly varying amplitude, $\Phi(r)$ is a rapidly varying phase,
\begin{equation}
k_r=\frac{d\Phi}{dr}=\frac{\sqrt{l(l+1)}}{r}\,\frac{N(r)}{\omega}\label{kr}
\end{equation}
is the radial wavenumber, and $N(r)$ is the Brunt-V\"ais\"al\"a
frequency.
One can see that due to the high value of $N/\omega$, the
following strong inequalities take place:
\begin{equation}
k_r\gg\frac{\sqrt{l(l+1)}}{r}\equiv k_H,\quad
|\mbox{\boldmath $\xi$}_H|\gg|\xi_r|
\quad(\mbox{except near}\ \sin\Phi=0). \label{ineq}
\end{equation}

Standard estimate of nonlinearity, i.e. comparing the $\partial
\mbox{\boldmath $v$}/\partial t$ and (\mbox{\boldmath
$v\cdot\nabla$})\mbox{\boldmath $v$} terms in the momentum
equation, is correct providing that the curvature effect is included.
Then we obtain
\begin{equation}
SN\equiv \max(|\mbox{\boldmath $\xi$}_H| k_r)\gtrsim 1\label{SN}
\end{equation}
as the criterion for the strong nonlinearity. This is different
from that given by Kumar and Goodman \shortcite{Kumar}. The most
precise way to obtain our criterion is to apply the asymptotic
approximation (Eqs. \ref{xi},\ref{kr}) to the amplitude expansion
of the Hamiltonian. However, at least fourth order expansion is
needed (see Van~Hoolst, \citeyear{VanHoolst}, who considers the
general case of the nonradial stellar oscillation).

\section{Surface amplitude at the onset of the strong nonlinearity}

As an application we considered the
$\ell=1$ and $2$ modes in three models of
evolved stars: a TAMS $\delta$-Scuti star, a post-MS $\delta$-Scuti star, and
an RR~Lyrae star. The two $\delta$-Scuti models have $\log T_{\rm eff}=3.86$
and the RR~Lyrae model has $\log T_{\rm eff}=3.84$.
\begin{figure}
\centerline{\includegraphics[height=0.8\textwidth,width=0.8\textwidth]
{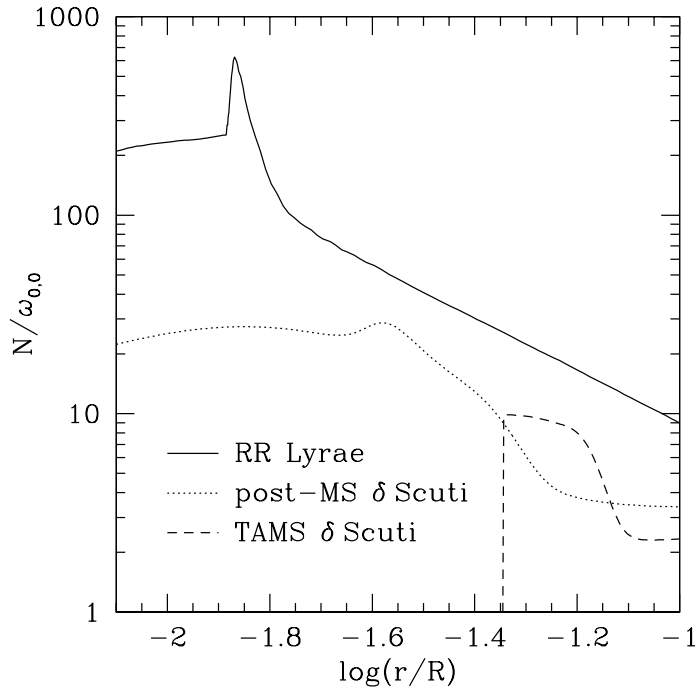}}
\caption{Ratio of the Brunt-V\"ais\"al\"a frequency
to the fundamental radial mode frequency in three selected models.}
\end{figure}

Fig.1 shows the ratio of the Brunt-V\"ais\"al\"a frequency to
the fundamental radial mode frequency in our models. The differences are
striking. The maximum value of this ratio in the RR~Lyrae model is at least
by an order of magnitude larger than those in $\delta$-Scuti models.
This suggests that nonlinearity in the interior may be a greater problem in
more evolved stars. To assess the problem we need an estimate of
\mbox{\boldmath $\xi$}$_H$ from
the observed surface amplitude. To this aim
we performed LNA calculations for the selected three models focusing on
linearly unstable modes of $\ell=1$ and $2$.
In this way we derived bolometric light amplitudes corresponding
to $SN=1$ which we call the critical amplitudes.

The
critical amplitudes for the TAMS star were found typically higher than 100
mmag which is
well above the values observed in MS $\delta$-Scuti stars.
Thus the nonlinearity is not a problem in this case.
For the post-MS model the critical amplitudes are much
lower, often below 10 mmag.
Some of the modes in 4CVn have observed amplitudes
exceeding the critical values.
Thus the nonlinearity in the deep interior may be a problem in this case.

In the case of RR~Lyrae model, critical amplitudes, shown in
Fig.2, are even lower. The highest values are found for $\ell=1$
modes in the vicinity of the first overtone. But even in this case,
the value of 0.02 mag is significantly lower than observed. We see
that the critical amplitudes are much lower near the fundamental
mode frequency and for the whole $\ell=2$ sequence. It is thus
clear that the observed close peaks in RR~Lyrae stars cannot be
interpreted in terms of linear nonradial eigenmodes.

The nonlinearity presents a problem for all models involving
nonradial motion, whether it is due to a nonradial mode excitation
or to a magnetically induced asphericity. A fully nonlinear
treatment of the motion is needed if we want to model the observed
amplitude modulation.

\begin{figure}
\centerline{\includegraphics[height=0.72\textwidth,width=0.8\textwidth]
{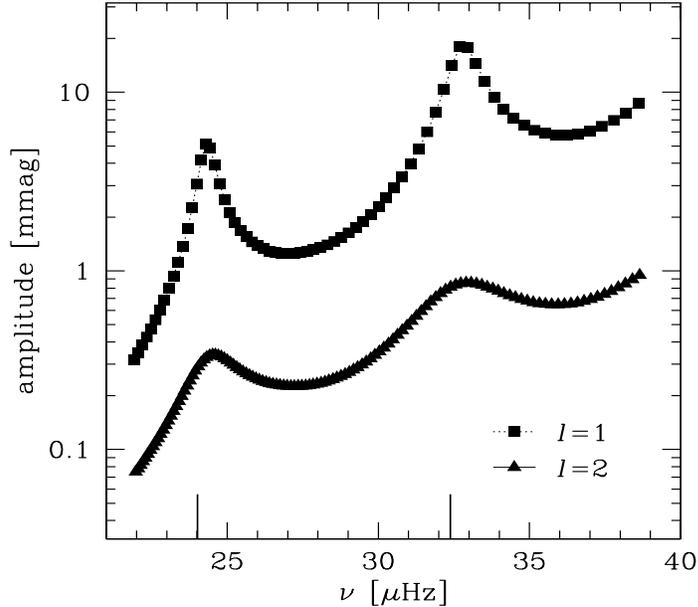}}
\caption{Critical amplitudes of the bolometric flux for nonradial modes in the
RR~Lyrae model. The long tickmarks denote frequencies of the lowest radial
modes.}
\end{figure}

Our work is supported by the KBN grant No. 5 P03D 030 20.


\end{article}
\end{document}